\documentclass[aps,prd,twocolumn,nofootinbib,superscriptaddress]{revtex4-2}
\usepackage{amsmath,amssymb,amsfonts}
\usepackage{graphicx}
\usepackage{bm}
\usepackage{hyperref}
\usepackage{color}
\begin{document}

\title{Dynamics of ($Z_N$) Domain Walls in SU(N) Gauge Theories}

\author{Sayanjit Banerjee}
\email{sayanjitb@imsc.res.in}
\author{Sanatan Digal}
\email{digal@imsc.res.in}
\author{Sumit Shaw}
\email{sumitshaw@imsc.res.in}
\affiliation{Institute of Mathematical Sciences, Chennai 600113, India\\ Homi Bhabha National Institute, Training School Complex, Anushaktinagar, Mumbai 400094, India}
\date{\today}

\begin{abstract}

We study collisions of domain walls in $SU(N)$ gauge theories using the Polyakov-loop effective potential models. We find that string junctions play a crucial role in the dynamics of $Z_N$ domain walls. In $SU(3)$ gauge theory, the merger of two non-planar $Z_3$ domain walls into a single wall proceeds via the creation of a vortex--antivortex pair in $2+1$ dimensions. In $SU(4)$ gauge theory, low-energy collisions of $Z_4$ walls result in the formation of a single domain wall without the creation of vortices.  At higher collision energies, two $Z_4$ domain walls can either bounce back or scatter into another pair of domain walls through the formation of vortices. The creation of vortex--antivortex pairs generalises to the creation of string loops in $3+1$ dimensions for both gauge theories. These results demonstrate a direct dynamical role for topological strings in the evolution of center-domain-wall networks and reveal a new aspect of defect dynamics in non-Abelian gauge theories.

 \end{abstract}


\maketitle

\section{Introduction}
Topological defects play a central role across a wide range of physical systems, from condensed-matter systems to high-energy quantum field theories and cosmological models of the early Universe. They often arise during phase transitions involving spontaneous symmetry breaking, when the vacuum manifold becomes degenerate~\cite{Kibble:1976sj}. In such situations, nontrivial field configurations interpolating between distinct vacua can form, giving rise to defects such as domain walls, strings, and monopoles~\cite{Kibble:1980mv,Vilenkin:2000jqa}. The types of defects that can occur are determined by the topology of the vacuum manifold and the dimensionality of space~\cite{Mermin:1979zz}. Understanding the properties and dynamics of these defects is often essential for understanding the behavior of the system in its symmetry-broken phase, including phase ordering kinetics, defect-network evolution, and nonequilibrium dynamics~\cite{Bray:1994zz,Hindmarsh:1994re}.

Finite-temperature pure SU(N) gauge theories provide an important example of a system with a rich topological structure. These theories undergo a confinement–deconfinement transition at high temperatures, with the Polyakov loop serving as an order parameter for the transition~\cite{Polyakov:1978vu,Svetitsky:1982gs,Yaffe:1982qf}. Above a critical temperature $T_c$, the Polyakov loop acquires a nonzero expectation value and the global center symmetry, $Z_N$, is spontaneously broken, leading to N degenerate deconfined vacua related by center transformations~\cite{tHooft:1977nqb,Svetitsky:1982gs}. Consequently, the deconfined phase supports domain walls separating distinct deconfined vacua~\cite{Bhattacharya:1992qb}.

Recently, we demonstrated that the deconfined phase of SU(N) gauge theories also supports topological string defects~\cite{tgs}. These strings arise as junctions of center domain walls and are characterized by nontrivial winding of the Polyakov-loop order parameter around their cores~\cite{Bhattacharya:1992qb,Korthals-Altes:1999cqo,tgs}. Similar wall-junction configurations have previously appeared in supersymmetric gauge theories~\cite{Dvali:1996xe,Kovner:1997ca}. The existence of such defects enriches the topological structure of the deconfined phase and suggests a close connection between string and domain-wall dynamics. In particular, since domain walls can terminate on or merge through string junctions, these defects provide a natural mechanism through which domain walls can interact and reorganize their topology.

The static properties of the domain walls in SU(N) gauge theories have been studied extensively using perturbative analyses, effective descriptions and lattice simulations \cite{Bhattacharya:1992qb,Korthals-Altes:1999cqo,deForcrand:2005pb,	deForcrand:2005rg,deForcrand:2004jt,Bursa:2005yv}. Domain-wall tensions, interface profiles and scaling properties are now reasonably well understood \cite{Bhattacharya:1992qb,Bursa:2005yv,West:1996sb,West:1996ej,Korthals_Altes:1995}. In particular, studies of interface tensions suggest attractive interactions between elementary walls and wetting phenomena near the deconfinement transition \cite{Frei:1989es,Karsch:1990yh,Huang:1990jf,Aoki:1993dm}. Together, these investigations have yielded important insights into the nonperturbative structure of finite-temperature gauge theories.

Despite the considerable progress in understanding the static properties of domain walls and the recent identification of topological string defects \cite{Bhattacharya:1992qb,Korthals-Altes:1999cqo,Bursa:2005yv,tgs}, much less is known about their interaction dynamics. Although existing studies suggest attractive interactions between elementary domain walls \cite{Frei:1989es,Karsch:1990yh,Aoki:1993dm}, the free-energy landscape governing their approach has not been investigated in detail. To our knowledge, there are no quantitative studies of how the free energy depends on the separation between domain walls, nor of the processes that occur when domain walls collide. Since collisions can naturally lead to the formation of string junctions and other composite topological configurations \cite{tgs,Dvali:1996xe,Kovner:1997ca}, understanding these processes is essential for developing a complete picture of defect dynamics in the deconfined phase.

In this work, we investigate interactions between center domain walls in effective Polyakov-loop theories describing the deconfined phase of SU($N$) gauge theories. In particular, we focus on the SU(3) and SU(4) cases. Using real-time numerical simulations in $2+1$ dimensions in the Polyakov loop effective potential models, we uncover a common and previously unexplored mechanism governing domain-wall interactions. Although our simulations are performed in two spatial dimensions, the mechanism extends naturally to $3+1$ dimensions, and we discuss the corresponding interaction processes in $3+1$ dimensions.

Our results show that, in SU(3), the collision of two domain walls dynamically generates a localized vortex--antivortex pair. As the vortex and antivortex separate, a third domain wall forms between them. A straightforward generalization to three spatial dimensions suggests that, instead of a vortex--antivortex pair, a string loop is created, forming the boundary of the third domain. Thus, the collision proceeds through the creation of the same topological string defect previously identified in lattice simulations of the full gauge theory~\cite{tgs}. For low collision energies, the domain walls merge to form a single domain wall through the vortex--antivortex-mediated process described above. For a different choice of parameters of the Polyakov-loop effective potential and sufficiently high collision energies, the wall formed in the collision subsequently splits back into the original pair through the creation of vortices in two dimensions and string loops in three dimensions.

The dynamics in SU(4) share some qualitative features with SU(3), but also display important differences. For low collision energies, two domain walls interpolating between nearest-neighbor vacua merge to form a single domain wall connecting next-to-nearest-neighbor vacua in the vacuum manifold. In this process, no vortices or string loops are created. At higher collision energies, however, the walls scatter into a different pair of nearest-neighbor domain walls. This process is mediated by the creation of vortices in two dimensions and string loops in three dimensions. These results indicate that string-mediated wall recombination and scattering are generic features of deconfined SU($N$) gauge theories.

The paper is organized as follows. In Sec.~II, we review the Polyakov loop effective potential model that describes the CD transition and spontaneous breaking of $Z_N$ in SU($N$) gauge theories. In Sec.~III, we discuss topological defects associated with spontaneous breaking of $Z_N$ symmetry; in particular, we focus on SU($3$) and SU($4$) gauge theories and obtain their spatial profiles using numerical simulations. Section~IV presents our results for the interaction of domain walls via strings for both SU($3$) and SU($4$). Finally, Sec.~V contains our conclusions and outlook.

\section{Effective Polyakov-Loop Model for $SU(N)$ Gauge Theories}

\subsection{Center symmetry and the Polyakov loop}

In pure $SU(N)$ gauge theories, the confinement--deconfinement (CD) phase transition is associated with the realization of the global center symmetry $Z_N$. The Polyakov loop serves as an order parameter for this transition~\cite{Polyakov:1978vu}. It is defined as,
\begin{equation}
L(\mathbf{x})=
\frac{1}{N}\,
\mathrm{Tr}\,
\mathcal{P}
\exp\!\left(
ig\int_0^{1/T}
A_0(\mathbf{x},\tau)\,d\tau
\right),
\label{eq:polyakov}
\end{equation}
 where $T$ denotes the temperature, $A_0$ is the temporal gauge field, and $\mathcal{P}$ represents path ordering in Euclidean time.

Under a gauge transformation that is periodic only up to an element of the center, $Z_N$, of $SU(N)$~\cite{tHooft:1977nqb,Polyakov:1978vu}, i.e
\begin{eqnarray}
&&A_\mu(x)\longrightarrow U A_\mu(x) U^{-1} + {i \over g}\left( \partial_\mu U\right)U^{-1},~~~\nonumber\\
&&U(\vec{x},\tau=0)=z^{-1}U(\vec{x},\tau=\beta),\nonumber\\
&&z=e^{2\pi i k/N}\in Z_N,~{\rm and}~k=0,\ldots,N-1,~~~~~~~~~~\nonumber
\end{eqnarray}
the Polyakov loop transforms as
\begin{equation}
L(\mathbf{x})
\rightarrow
z\,L(\mathbf{x}).
\end{equation}
In the confined phase, 
\begin{equation}
\langle L\rangle =0,
\end{equation}
thus, the center symmetry remains unbroken~\cite{Polyakov:1978vu,tHooft:1977nqb}. Above the critical temperature $T_c$, the Polyakov loop acquires a non-vanishing expectation value,
\begin{equation}
\langle L\rangle \neq 0.
\end{equation}
Consequently, the expectation value of the Polyakov loop transforms non-trivially under the center symmetry, signalling spontaneous breaking of the $Z_N$ symmetry~\cite{Svetitsky:1982gs,Yaffe:1982qf}. The deconfined phase therefore contains $N$ degenerate vacua related by center transformations~\cite{Svetitsky:1982gs,Yaffe:1982qf,tHooft:1977nqb}. The emergence of these degenerate vacua plays a central role in the formation of topological defects in the deconfined phase.


\subsection{The Polyakov-loop effective potential}

Near the confinement--deconfinement transition, the long-distance dynamics of the SU($N$) theory can be described in terms of the Polyakov loop effective potential for the complex order parameter $L$, which is similar to Landau--Ginzburg models~\cite{Pisarski:2000eq,Dumitru:2000in,Svetitsky:1982gs}. The effective potential is constrained by the underlying $Z_N$ center symmetry. Polyakov-loop effective theories have also been extended to include dynamical quarks and have found widespread application in studies of the QCD phase structure and dynamics~\cite{Fukushima:2003fw,Ratti:2005jh}.

For $SU(3)$, the effective potential consistent with $Z_3$ symmetry is
\begin{equation}
V(L) = {b_2 \over 2}\,|L|^2 - {b_3 \over 3} \left(L^3+L^{*3}\right) + {b_4\over 4}\,|L|^4,~~ L=L_R+iL_I,
\label{z3}
\end{equation}
where $L_R$ and $L_I$ denote the real and imaginary components of the Polyakov-loop field, respectively.
Choosing $b_2$ to be temperature dependent, while treating $b_3$ and $b_4$ as constants, yields a good description of the thermodynamics of pure $SU(3)$ gauge theory as determined from non-perturbative lattice simulations~\cite{Pisarski:2000eq,Dumitru:2000in,Svetitsky:1982gs}. The cubic term, which is allowed by the $Z_3$ symmetry, is also responsible for the first-order nature of the confinement--deconfinement transition in pure $SU(3)$ gauge theory. For $T> T_c$, in the deconfined phase, the potential develops three degenerate minima related by $Z_3$ transformations.

The above form can be generalized to higher-rank gauge groups as
\begin{equation}
V(L) = a_N|L|^2 + c_N|L|^4 + d_N\left(L^N+L^{*N}\right) +\cdots ,
\label{gen}
\end{equation}
for SU($N$). Note that higher-order $Z_N$-invariant operators,
in Eq.\ref{gen}, may be required to ensure stability of the potential and to reproduce the correct thermodynamic properties of the underlying gauge theory. In the deconfined phase, the minima of the effective potential correspond to the distinct center-related vacua.

\begin{figure}[t]
	\centering
	\includegraphics[width=0.50\linewidth]{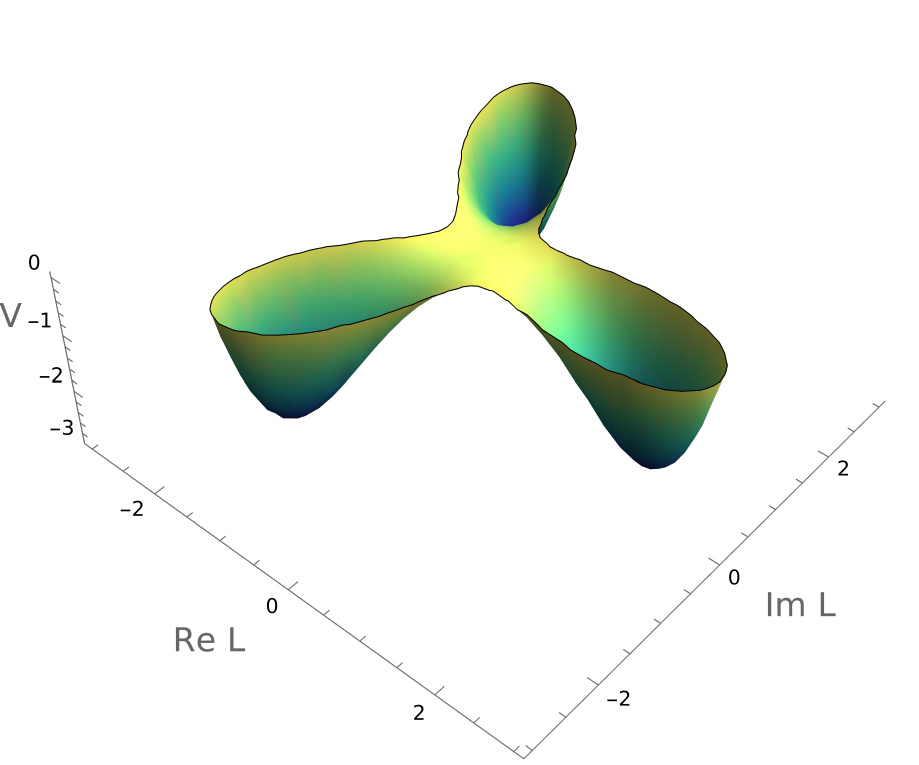}
	\caption{The Polyakov loop effective potential $V(L)$ in complex-L plane exhibiting spontaneous breaking of $Z_3$ symmetry with three degenerate minima.}
	\label{pot1}
\end{figure}

\begin{figure}[t]
	\centering
	\includegraphics[width=0.50\linewidth]{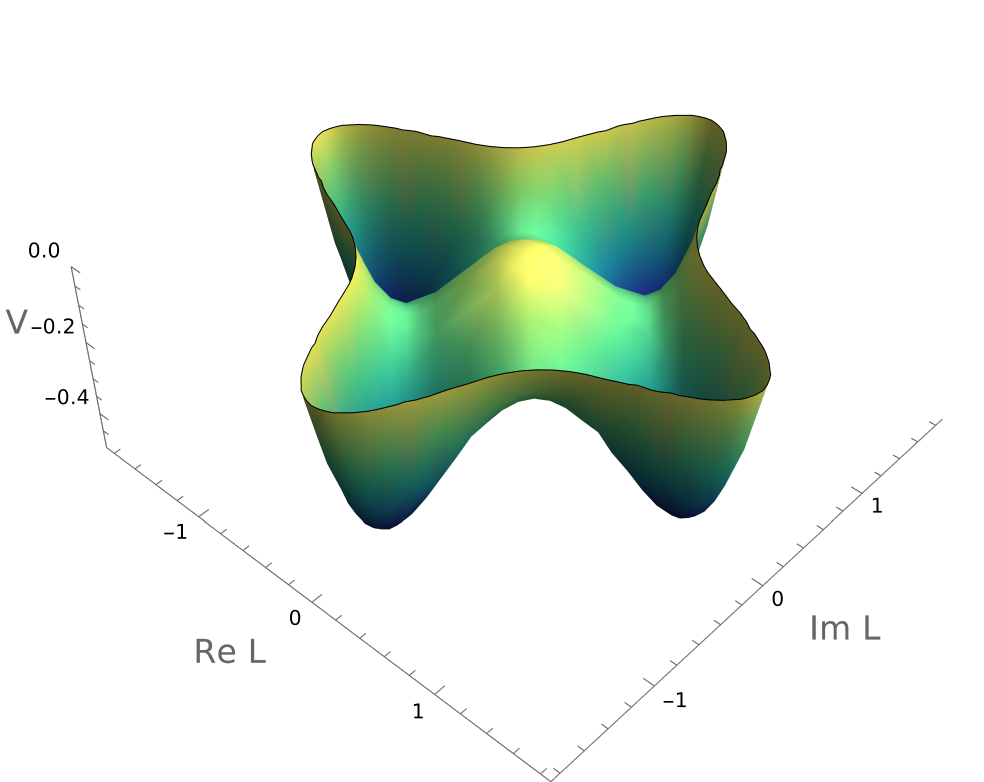}
	\caption{The Polyakov loop effective potential $V(L)$ in complex-L plane exhibiting spontaneous breaking of $Z_4$ symmetry with four degenerate minima.}
	\label{pot2}
\end{figure}

In the deconfined phase, typical forms of the effective potential for $SU(3)$ and $SU(4)$ are shown in Figs.~\ref{pot1} and \ref{pot2}, respectively. In both cases, the center symmetry is spontaneously broken.
In the deconfined phase of $SU(3)$ gauge theory, spontaneous breaking of the $Z_3$ symmetry leads to three degenerate vacua,
\begin{equation}
L = \eta\left(1,e^{2\pi i/3},e^{4\pi i/3}\right)\equiv(L_1,L_2,L_3),
\end{equation}
where $\eta=|\langle L\rangle|$ denotes the magnitude of the Polyakov-loop expectation value. In the deconfined phase of $SU(4)$ gauge theory, spontaneous breaking of the $Z_4$ symmetry leads to four degenerate vacua,
\begin{equation}
L=\eta\left(1,e^{i\pi/2},e^{i\pi},e^{3i\pi/2}\right)\equiv(L_1,L_2,L_3,L_4),
\end{equation}
where $\eta=|\langle L\rangle|$ is the magnitude of the Polyakov-loop.


\section{Topological Defects Associated with Broken $Z_N$ Symmetry}

To study the long-distance properties and spacetime evolution of topological defects associated with the spontaneously broken center symmetry, we employ a Landau--Ginzburg effective action for the Polyakov-loop field,
\begin{equation}
S_{\rm eff}
=
\int d^4x
\left[
K\,\partial_\mu L^*\partial^\mu L
-
V(L)
\right],
\label{frn}
\end{equation}
where $K$ is a temperature-dependent coefficient~\cite{Pisarski:2000eq,Dumitru:2000in,Scavenius:2002ru}. The kinetic term governs the spacetime variations of the Polyakov-loop field, while the effective potential determines the vacuum structure associated with the center symmetry.  

\subsection{Domain walls and strings in $SU(3)$ gauge theory}

In the spontaneously broken $Z_3$ phase, shown in Fig.~\ref{pot1}, there are three degenerate vacua,
\[
L_1=\eta,\qquad
L_2=\eta e^{2\pi i/3},\qquad
L_3=\eta e^{4\pi i/3}.
\]
Consequently, there are three domain-wall solutions interpolating between pairs of vacua~\cite{Bhattacharya:1992qb}. We denote by $W_{ij}$ the domain wall connecting the vacua $L_i$ and $L_j$. Across each wall, the phase of the Polyakov loop changes by $2\pi/3$.

A $Z_3$ string is formed when all three domain walls meet along a common line~\cite{tgs}. As one traverses a closed loop encircling the string, the phase of the Polyakov loop winds by $2\pi$, passing successively through the three $Z_3$ vacua. Continuity of the field then requires the Polyakov loop to pass through the confining value $L=0$ at the center of the configuration. Consequently, the string core corresponds to a locally restored confining phase and forms the junction of the three domain walls. Such configurations are topologically stable within the effective Polyakov-loop theory, since the nontrivial winding of the order parameter cannot be continuously unwound without forcing the field through the confining point $L=0$.

To obtain the domain-wall and string configurations, we numerically solve the equations of motion for the Polyakov-loop field derived from the effective action, Eq.~(\ref{frn}), using the parameters $b_2=-0.27$, $b_3=2.0$, $b_4=1.0$, and $K=1.649$~\cite{Pisarski:2000eq,Dumitru:2000in,Scavenius:2002ru}. The resulting domain walls and strings correspond to static solutions of the field equations. However, instead of solving the static field equations, for simplicity we evolve a suitably chosen initial configuration using time-dependent field equations including a small dissipative term~\cite{Atreya:2014gua}. The field equations with dissipation coefficients are as follows,

\begin{eqnarray}
\label{feq}
&&2K\frac{\partial^2\phi_i}{\partial t^2}+\xi\frac{\partial\phi_i}{\partial t}-\nabla^2\phi_i=-\frac{\partial V}{\partial\phi_i},~ i=1,2, \\
&&{\rm where},~\nabla^2 = {d^2 \over dx^2}+{d^2 \over dy^2},\nonumber
\end{eqnarray}

and $\xi$ is a damping coefficient. The dissipative term removes excess energy from the system while preserving the underlying vacuum structure. As the evolution proceeds, the field velocities and accelerations approach zero, and the configuration relaxes to a stationary defect solution corresponding to a local minimum of the effective action. This relaxation procedure provides an efficient method for obtaining domain-wall and string configurations. The same field equations can also be used to investigate the subsequent dynamical evolution and interactions of these defects.

\begin{figure}[t]
	\centering
	\includegraphics[width=0.750\linewidth]{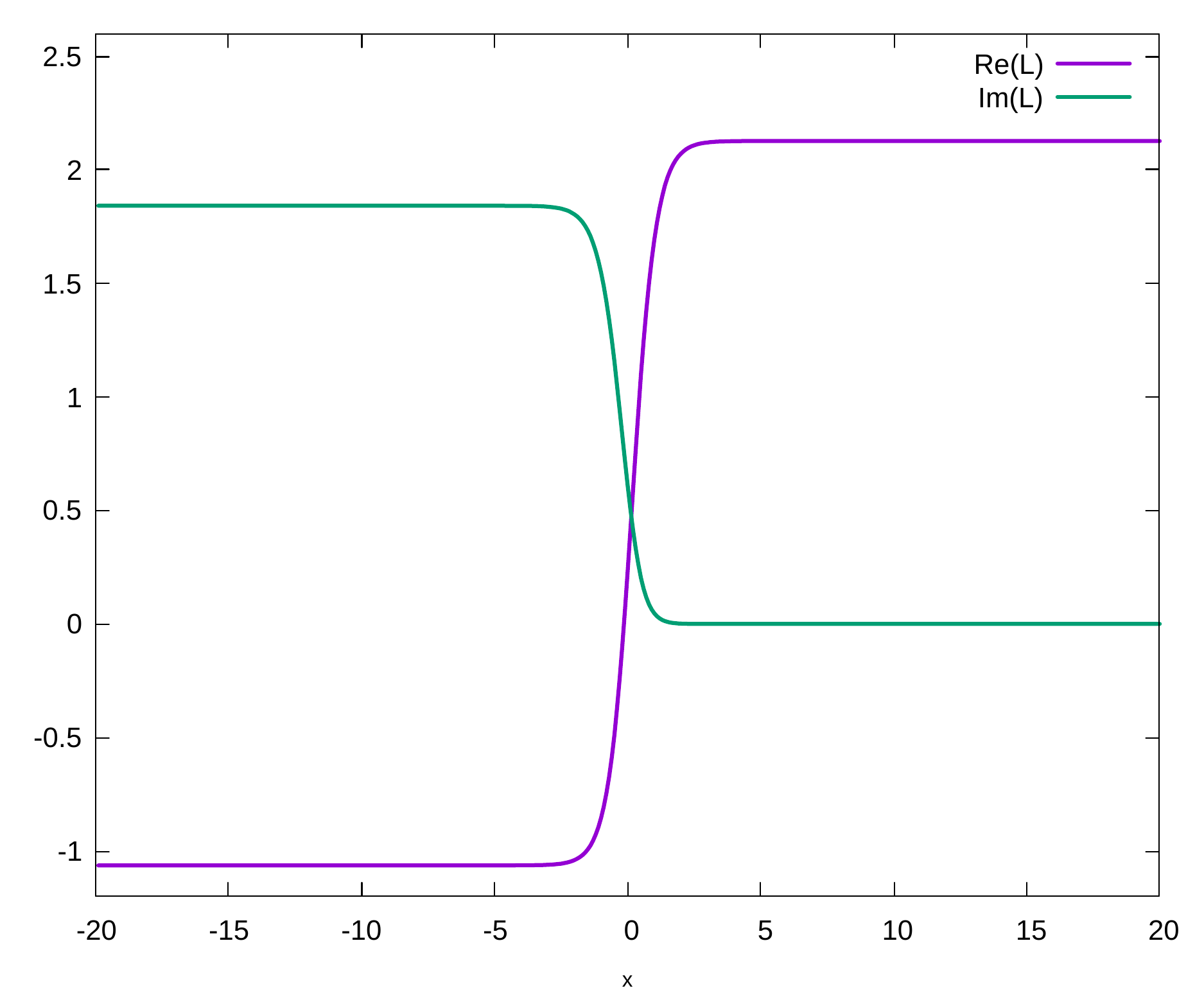}
	\caption{The real and imaginary components of the stationary $W_{12}$ domain-wall profile in the broken $Z_{3}$ phase. The Polyakov-loop field interpolates from $L_{1}$ to $L_{2}$, following dissipative relaxation under Eq.~\ref{feq} with $Z_3$ potential.}
	\label{z3dw}
\end{figure}

For the $W_{12}$ domain-wall solution, we impose the boundary conditions
\begin{equation}
\lim_{x\rightarrow -R/2}L=L_1,
\qquad
\lim_{x\rightarrow +R/2}L=L_2,
\end{equation}
where $R$ denotes the spatial extent of the lattice and is chosen to be much larger than the expected wall thickness. Twisted boundary conditions are imposed along the spatial direction, which solves the problems that arise from boundary effects~\cite{Kajantie:1990bu}. Starting from an interpolating field configuration, the system is evolved using Eq.~(\ref{feq}) until the residual oscillations in the total energy become negligible. More precisely, the evolution is terminated when the variation in a single time step falls below $10^{-8}$, which is the numerical precision of the code. The real and imaginary parts of the Polyakov loop profile of the resulting $W_{12}$ domain-wall is shown in Fig.~\ref{z3dw}, interpolating between $L_1$ and $L_2$ vacua.

\par
\begin{figure}[t]
	\centering
	\includegraphics[width=0.50\linewidth]{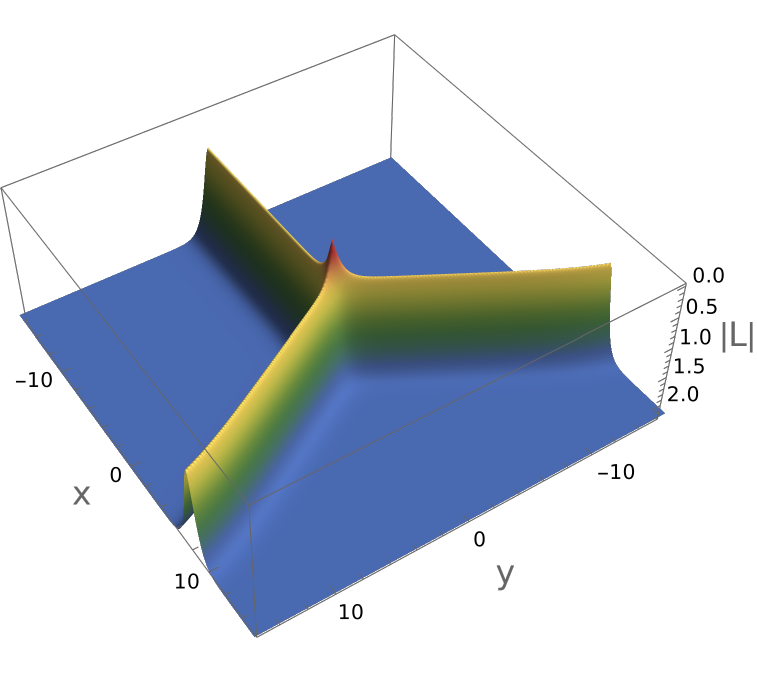}
	\caption{Spatial profile of the $Z_3$ string configuration formed at the junction of three domain walls. The string core corresponds to the confining phase where $L\simeq 0$.}
	\label{z3s}
\end{figure}

 Since the minimum-energy configuration of an isolated string is translationally invariant along its length, it is sufficient to determine the field configuration in a plane perpendicular to the string. The resulting field configuration is effectively that of a vortex (or antivortex) in two spatial dimensions. The initial configuration is therefore chosen to possess the appropriate winding structure around the center of the lattice. We choose an appropriate configuration at the boundary of the lattice that is consistent with a string configuration at the center. Upon relaxation according to Eq.~\ref{feq}, the configuration evolves into a stable $Z_3$ string connected to three domain walls. The resulting string solution is shown in Fig.~\ref{z3s}. Here, the $Z_3$ string is formed at the junction of three domain walls associated with the three degenerate $Z_3$ vacua.

\subsection{Domain walls and strings in $SU(4)$ gauge theory}

The spontaneously broken $Z_4$ phase contains four degenerate vacua,
\[
L_1=\eta,\quad
L_2=\eta e^{i\pi/2},\quad
L_3=\eta e^{i\pi},\quad
L_4=\eta e^{i3\pi/2}
\]
related by center transformations. The corresponding vacuum structure supports both domain walls and string-like defects~\cite{Bhattacharya:1992qb,Korthals-Altes:1999cqo}. The construction of these defects is analogous to the $SU(3)$ case. A nontrivial winding of the Polyakov loop around a closed contour forces the field to pass through the confining value L=0 at the defect core, giving rise to a topologically stable Z4 string.

The four vacua give rise to six distinct domain-wall solutions. We denote by $W_{ij}$ the wall interpolating between the vacua $L_i$ and $L_j$. Four of these walls connect adjacent vacua in the complex-$L$ plane,
\[
W_{12}\equiv (L_1,L_2),\quad
W_{23}\equiv (L_2,L_3),\quad
\]
\[
W_{34}\equiv (L_3,L_4),\quad
W_{14}\equiv (L_4,L_1),
\]
while the remaining two connect antipodal vacua,
\[
W_{13}\equiv (L_1,L_3),\qquad
W_{24}\equiv (L_2,L_4).
\]
Unlike the $SU(3)$ case, the presence of both adjacent-vacuum and antipodal walls introduces multiple channels for wall recombination and scattering. As we will show in Sec.~IV, this richer defect spectrum leads to qualitatively new collision dynamics.

Stationary wall and string configurations are obtained using the same relaxation procedure described for the $SU(3)$ theory. For this purpose, we use the parameter values $a_4=-0.25$, $c_4=0.4$, and $d_4=-0.1$ in the effective potential, Eq.~\ref{gen}. The coefficient $K$ in the effective action, Eq.~\ref{frn}, is taken to be $K=1$. The Polyakov-loop field is evolved according to Eq.~\ref{feq}, with a small dissipative term, until a stationary configuration corresponding to a local minimum of the effective action is reached.

\begin{figure}[t]
	\centering
	\includegraphics[width=0.50\linewidth]{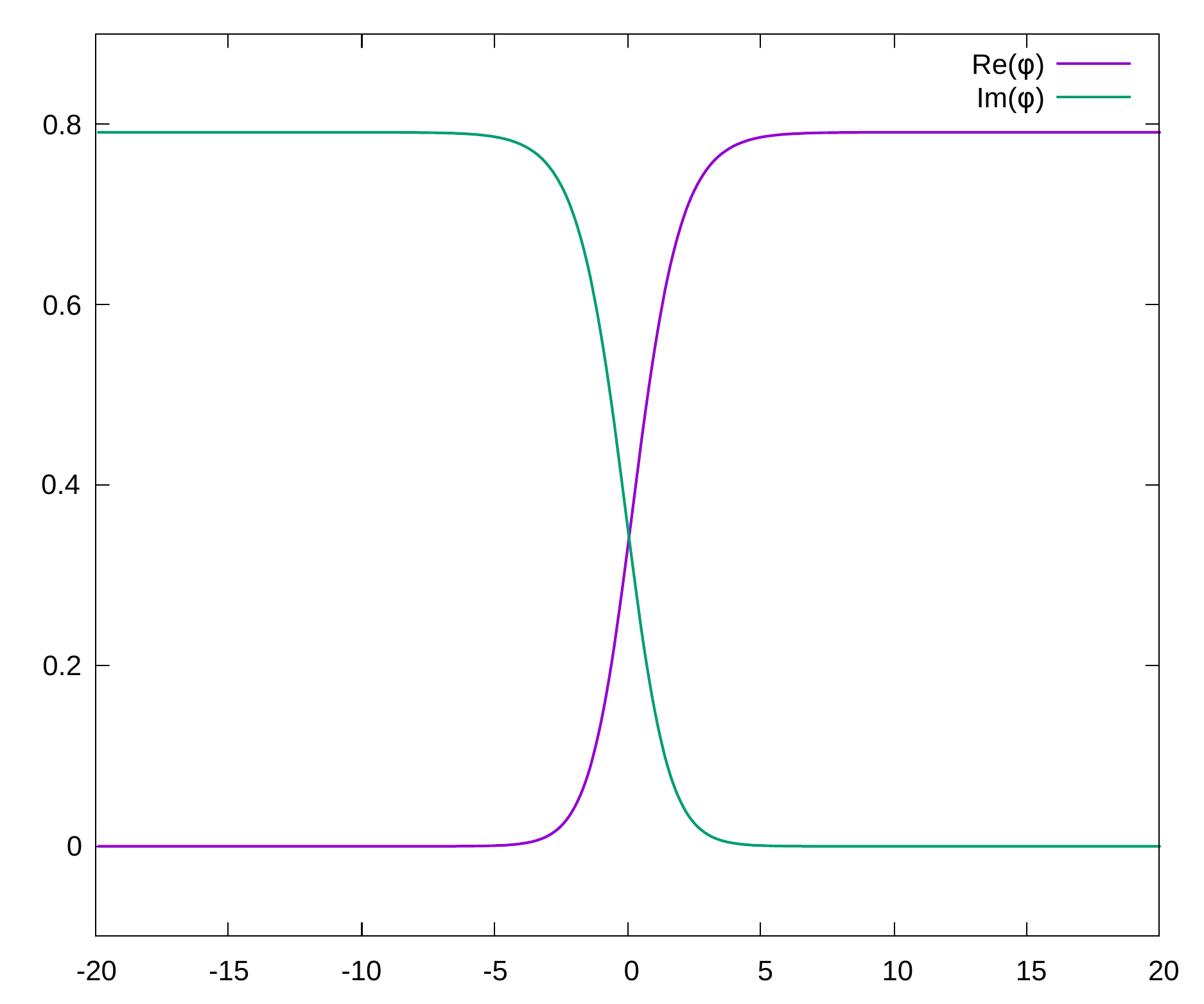}
	\caption{The real and imaginary components of the stationary $W_{12}$ domain-wall profile in the broken $Z_{4}$ phase. The Polyakov-loop field interpolates between adjacent vacua $L_{1}$ and $L_{2}$, following dissipative relaxation under Eq.~\ref{feq} with $Z_4$ potential.}
	\label{z4dw}
\end{figure}

To obtain the domain-wall solutions, we impose boundary conditions corresponding to the desired pair of vacua. As an example, the $W_{12}$ wall connecting the vacua $L_1$ and $L_2$ is obtained by imposing
\begin{equation}
\lim_{x\rightarrow -R/2}L=L_1,
\qquad
\lim_{x\rightarrow +R/2}L=L_2,
\end{equation}
where $R$ denotes the spatial extent of the lattice and is chosen to be much larger than the expected wall thickness. Twisted boundary conditions are imposed along the spatial direction. Starting from an interpolating field configuration, the system is evolved according to Eq.~\ref{feq}, until the residual oscillations in the total energy become negligible. The evolution is terminated when the amplitude of the energy oscillations falls below $10^{-8}$, as previously mentioned. The real and imaginary part of the Polyakov loop, which interpolates between $L_1$ and $L_2$ vacua of the $Z_4$ broken potential, is shown in Fig.~\ref{z4dw}.

\begin{figure}[t]
	\centering
	\includegraphics[width=0.50\linewidth]{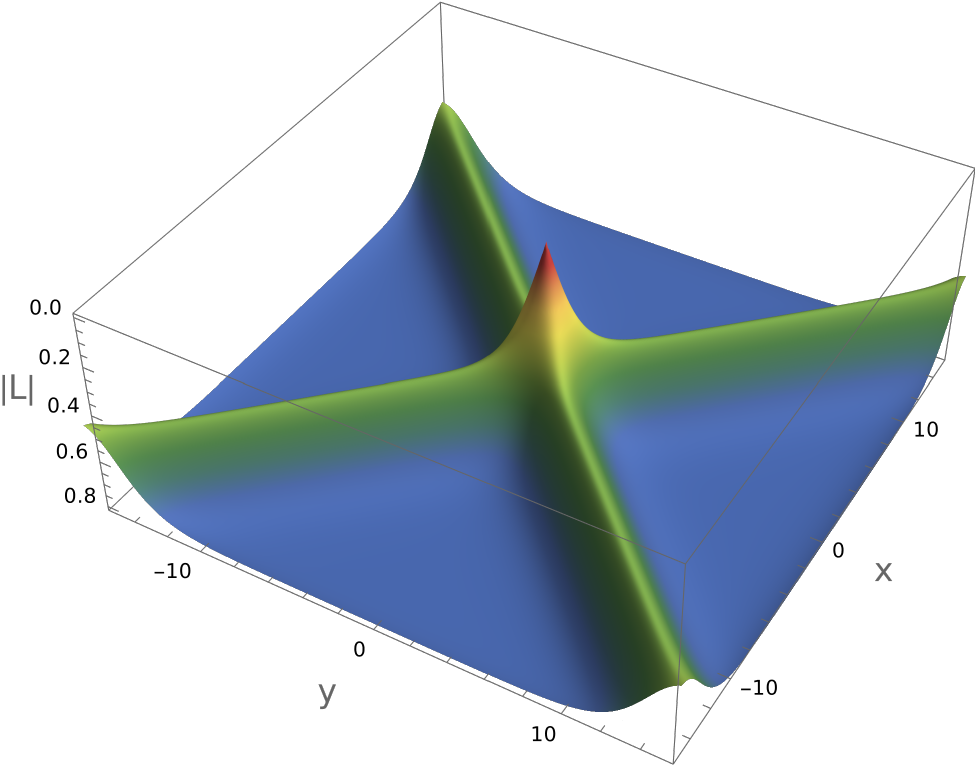}
	\caption{ Spatial profile of the $Z_4$ string configuration formed at the junction of four domain walls. The string core corresponds to the confining phase where $L\simeq 0$.}
	\label{z4s}
\end{figure}

The construction of the $Z_4$ string proceeds analogously to the $SU(3)$ case. An initial field configuration with the appropriate winding is evolved according to Eq.~\ref{feq}, until it relaxes to a stationary solution. The resulting configuration consists of a string core from which four domain walls emerge with approximate angular separations of $90^\circ$. No spontaneous rearrangement into a configuration involving antipodal walls is observed. Thus, the fundamental $Z_4$ string naturally appears as a junction of four adjacent-vacuum walls. The Fig.~\ref{z4s} shows the $Z_4$ string formed at the junction of four domain walls separating the four degenerate $Z_4$ vacua.

\section{Interaction of Domain Walls via Strings}

\subsection{Interactions of domain walls in $SU(3)$ gauge theory}

We now investigate collisions of domain walls in the effective Polyakov-loop model for SU(3) gauge theory. We begin with planar collisions between a $W_{12}$ wall and a $W_{23}$ wall in $(2+1)$ dimensions. Depending on the collision energy and the parameters of the effective potential, the collision can lead either to the formation of a $W_{13}$ wall or to a subsequent re-separation of the original walls.

\begin{figure}[t]
	\centering
	\includegraphics[width=0.65\linewidth]{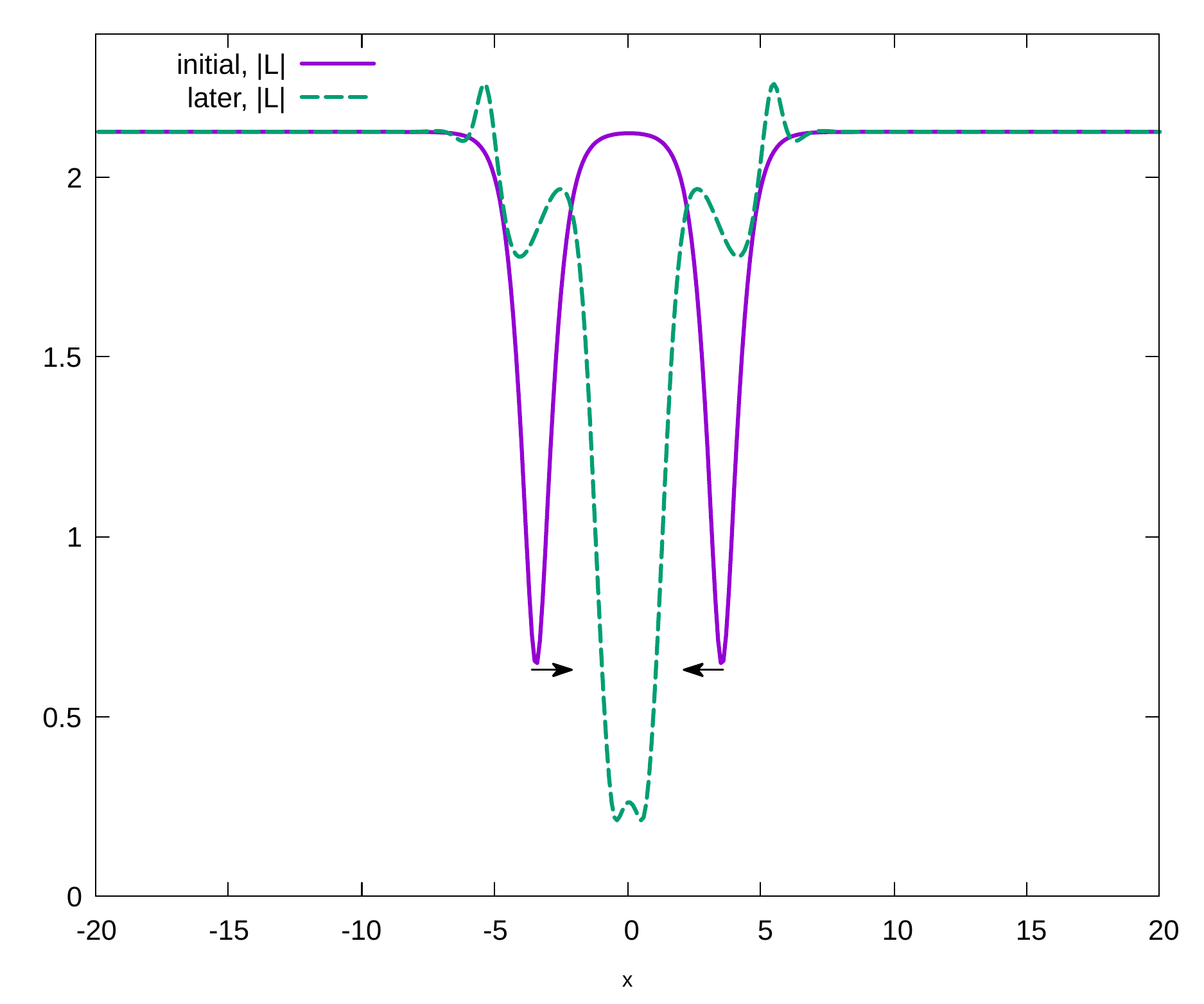}
	\caption{ Initial (solid line) and late-time (dashed) profiles of the Polyakov-loop magnitude during a low-energy planar collision of $Z_3$ walls ($W_{12}$ and $W_{23}$) in $(2+1)$ dimensions. The two incoming walls merge to form a single $W_{13}$ wall; the arrows indicate the directions of motion of the incoming walls.}
	\label{su3_planar_low}
\end{figure}

For the collision studies presented below, we use the same parameter values as those used in Sec.~III(A).
The initial configuration consists of a $W_{12}$ wall and a $W_{23}$ wall approaching one another. For low collision energies, the Polyakov-loop field in the interaction region evolves away from the intermediate vacuum $L_2$ toward the confining point $L=0$. It subsequently approaches the field value corresponding to the core of a $W_{13}$ wall, resulting in the merger of the two incoming walls into a single $W_{13}$ wall. Representative snapshots before and after the collision are shown in Fig.~\ref{su3_planar_low}. The solid curve shows the magnitude profile of the Polyakov loop for the two domain walls along the $x-$axis. The dashed curve represents the same for the configuration after the collision. 

The merger of $W_{12}$ and $W_{23}$ releases the binding as well as initial kinetic energy into both radial and phase excitations of the Polyakov-loop field, which propagate away from the collision region.
The radial modes remain localized in the collision region, as seen in Fig.~\ref{su3_planar_low}.  the collision energy is increased, the amplitudes of these excitations become larger and cover a larger area. Immediately after the collision, the field remains close to $L=0$ for some time. The field in the region does not flip back to the $L_2$ vacuum.

\begin{figure}[t]
	\centering
	\includegraphics[width=0.60\linewidth]{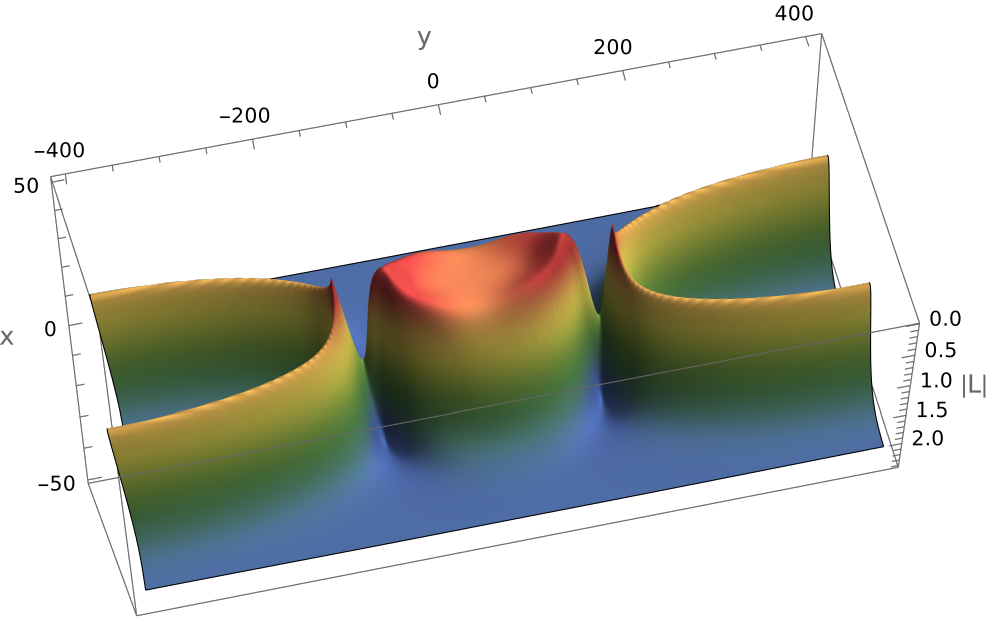}
	\caption{Collision of non-planar $W_{12}$ and $W_{23}$ domain walls resulting in formation of a vortex--antivortex pair and large fluctuations of the Polyakov loop in the $W_{13}$ wall, at zero damping. The vortices appear at the endpoints of the newly formed $W_{13}$ segment.}
	\label{su3_vortex_pair}
\end{figure}

\begin{figure}[t]
	\centering
	\includegraphics[width=0.60\linewidth]{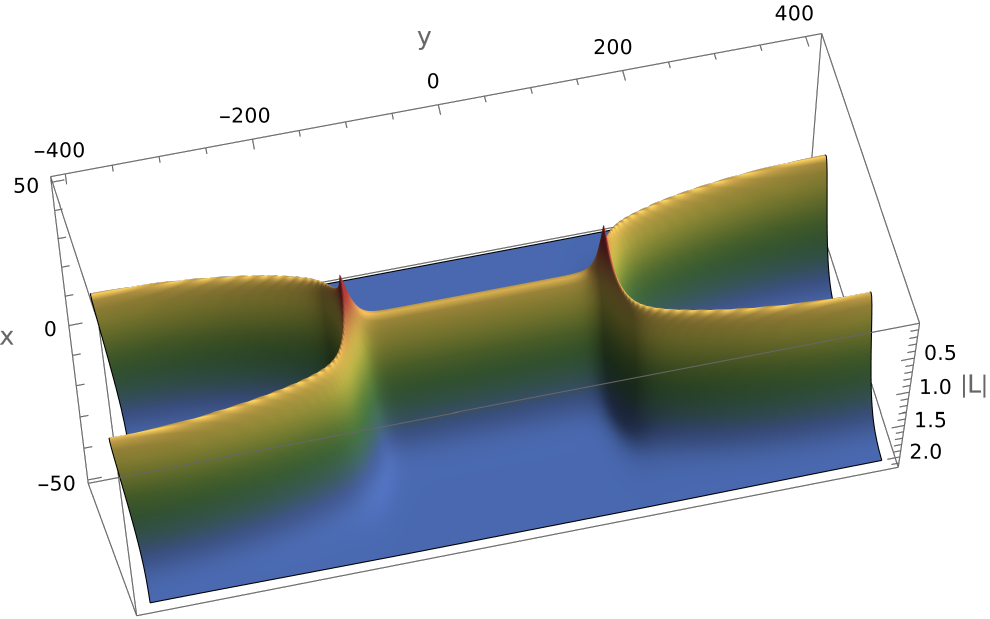}
	\caption{Formation of a vortex--antivortex pair in the presence of time-dependent damping, which suppresses fluctuations. The spatial profile of $|L(x,y)|$ shows that the formation of a vortex--antivortex pair appears at the endpoints of the newly formed wall segment.}

	\label{damping}
\end{figure}

More interesting phenomena arise in collisions of non-planar domain walls. In this case, different segments of the walls collide at different times. When local segments of the $W_{12}$ and $W_{23}$ walls merge, a corresponding segment of the $W_{13}$ wall is formed. The endpoints of the newly created $W_{13}$ segment are topologically nontrivial and correspond to a vortex--antivortex pair. The spatial profile of the Polyakov loop magnitude in the $xy-$ plane is shown in Fig.~\ref{su3_vortex_pair}, localized zeros of $|L|$ correspond to the vortex--antivortex winding, which can be understood from the vector plot of the real and imaginary components of the Polyakov loop of the same spatial magnitude profile. Thus, the conversion of $W_{12}$ and $W_{23}$ walls into a $W_{13}$ wall proceeds through the dynamical creation of topological defects. As the collision proceeds, the vortex and antivortex move apart along the collision front while additional portions of the incoming walls merge and are converted into the $W_{13}$ wall. The oscillations observed in planar collisions are also visible in the non-planar collisions along the newly formed wall segment in Fig.~\ref{su3_vortex_pair}.
With a time-dependent damping, the oscillatory excitations are suppressed, and the topological structure becomes manifest, as seen in Fig.~\ref{damping}. The vortex and antivortex appear as junctions of three domain walls, corresponding to the two-dimensional analogue of the $Z_3$ string discussed in Sec.~III.

At higher collision energies, the radial and phase excitations of the Polyakov loop become more prominent and cover a larger area. Though additional vortices form in these regions, they decay or annihilate before separating into fully formed vortices. For some choice of potential parameters, i.e $b_2=-0.87$, additional vortices do form; however, this value of $b_2$ is not consistent with pure $SU(3)$ gauge theory dynamics.



The vortex--antivortex pair creation observed in $(2+1)$ dimensions admits a natural interpretation in $(3+1)$ dimensions. In the higher-dimensional theory, vortices are replaced by $Z_3$ strings and antivortices by anti-strings. The transient splitting of the $W_{13}$ wall therefore corresponds to the nucleation of a closed string loop on the wall, followed by its subsequent contraction and annihilation. These processes reveal the direct dynamical role played by topological strings in domain-wall interactions.

\subsubsection{Free-energy landscape in deconfined $SU(3)$}

While the collision dynamics discussed above reveal the possible outcomes of domain-wall interactions, it is also useful to examine the free-energy landscape of a two-wall system. Although the separation between two walls becomes difficult to define once their profiles begin to overlap, the value of the Polyakov loop at the midpoint between the walls, denoted by $L_{\rm mid}$, evolves monotonically during the collision process. This suggests that $L_{\rm mid}$ may serve as a convenient parameter for characterizing overlapping two-wall configurations.

\begin{figure}[t]
	\centering
	\includegraphics[width=0.75\linewidth]{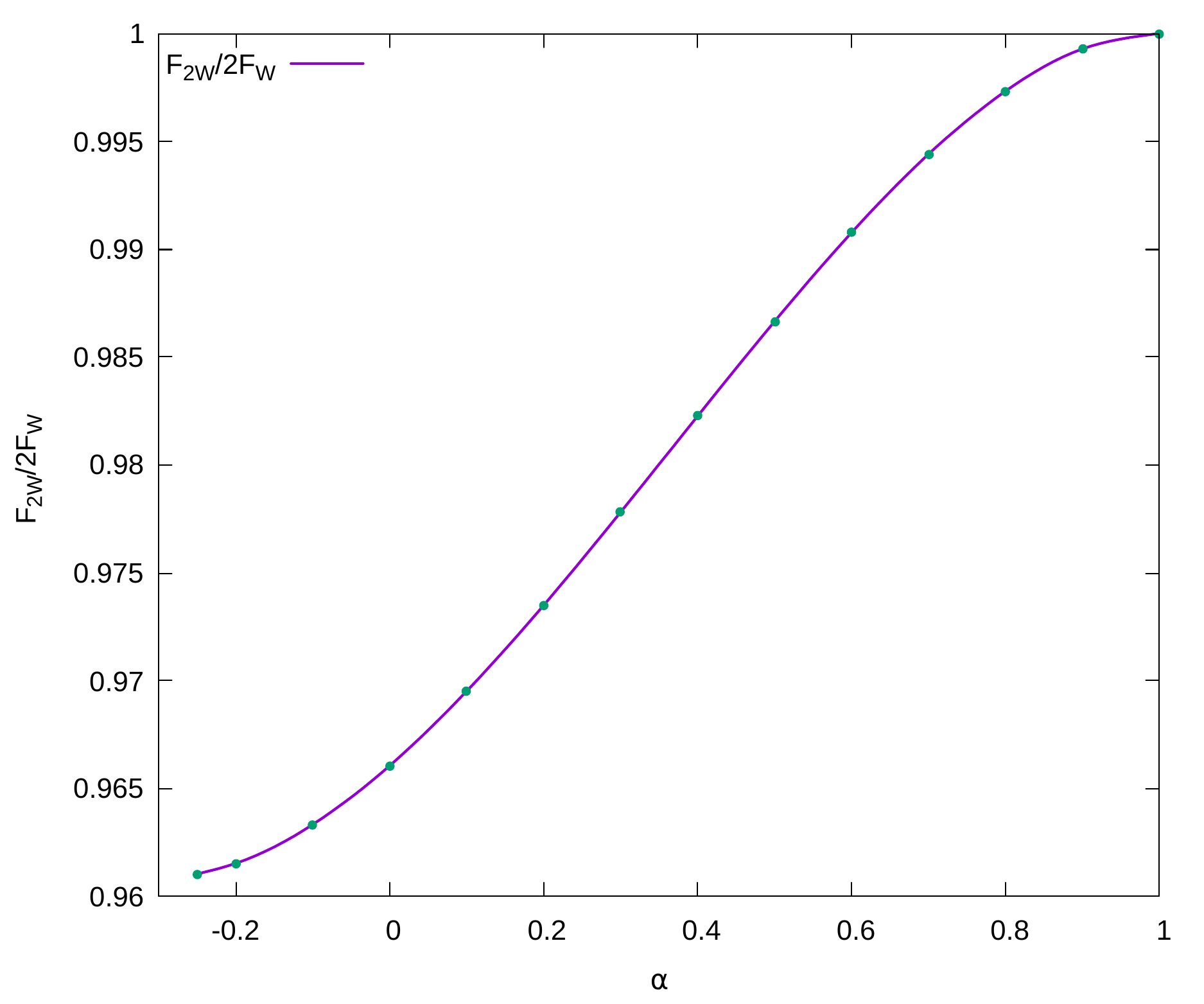}
	\caption{The normalized free energy, $F_{2W}/(2F_{W})$, of a constrained $W_{12}+W_{23}$ configuration in the $Z_{3}$ theory as a function of the midpoint coordinate $\alpha$. Here $\alpha=1$, $0$, and $-0.3$ correspond respectively to $L_{\mathrm{mid}}=L_{2}$, $L_{\mathrm{mid}}=0$, and the core value $L_{\mathrm{mid}}\simeq-0.3L_{2}$ of the resulting $W_{13}$ wall. The decrease of the free energy as the walls overlap demonstrates an attractive interaction.}
	\label{su3_fe}
\end{figure}

To determine the free energy($F_{2W}$) as a function of $L_{\rm mid}$, we consider a configuration consisting of a $W_{12}$ wall connecting the vacua $L_1$ and $L_2$ and a $W_{23}$ wall connecting the vacua $L_2$ and $L_3$. We fix the value of $L_{\rm mid}$ and evolve the field configuration using Eq.~\ref{feq}, for the effective action \ref{frn}, supplemented by a damping term to remove excess energy. Repeating this procedure for different values of $L_{\rm mid}$ along a straight-line path in the complex-$L$ plane connecting $L_2$ to the core value, $L_{core}\simeq-0.3L_2$, of the $W_{13}$ wall yields the normalized free-energy profile shown in Fig.~\ref{su3_fe}. In the figure, instead of plotting $F_{2W}$ we plot the ratio, $F_{2W}/2F_W$, where $F_W$ is the free energy of a single wall. $\alpha$ represents cordinate along a line connecting $L_2$ and the origin in the complex $L-$plane. $\alpha=1,0$, and $-0.3$ corresponds to $L_{mid}=L_2$, $(0,0)$, and $L_{core}$ respectively. The monotonic decrease of $F_{2W}$ in $\alpha$, indicates an attractive interaction between the $W_{12}$ and $W_{23}$ walls.

\subsection{Interactions of domain walls in $SU(4)$ gauge theory}

In the following, we consider domain-wall dynamics in the effective Polyakov-loop model for $SU(4)$ gauge theory.  For this purpose, we use the same parameters in the effective action, Eq.~\ref{frn}, as in Sec.~III(B). The presence of both adjacent-vacuum walls ($W_{12},W_{23},W_{34},W_{14}$) and antipodal walls ($W_{13},W_{24}$) gives rise to collision channels that are absent in $SU(3)$. We observe three qualitatively distinct outcomes for collisions between $W_{12}$ and $W_{23}$ walls, depending on the collision energy: (i) fusion into a $W_{13}$ wall, (ii) temporary fusion followed by bounce-back into the original wall pair, and (iii) transmutation into a different wall pair, $W_{14}+W_{43}$. A notable difference from the $SU(3)$ theory is that the low-energy fusion process does not require vortex--antivortex pair creation, even for non-planar collisions.

\begin{figure}[t]
	\centering
	\includegraphics[width=0.75\linewidth]{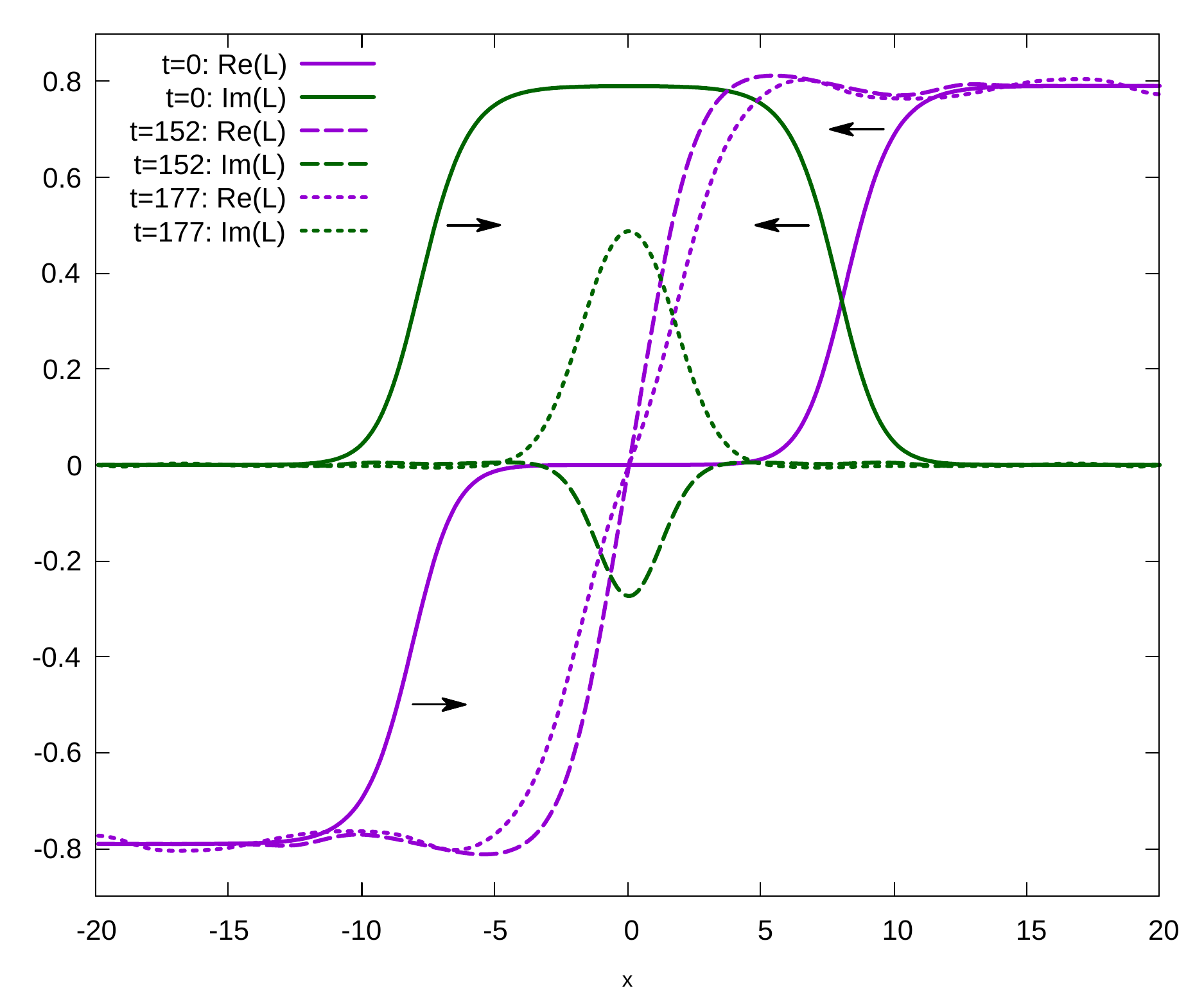}
	\caption{The evolution of the real and imaginary components of the Polyakov-loop field during a low-energy planar collision of $W_{12}$ and $W_{23}$ walls in the $Z_{4}$ theory. The initial (solid line) and late-time (dotted and dashed lines) profiles at the indicated times show the fusion of the two incoming walls into a single $W_{13}$ wall; arrows indicate the wall motion.}
	\label{su4_planar_low}
\end{figure}

For sufficiently low collision energies, the collision of the planar $W_{12}$ and $W_{23}$ walls leads to the formation of a single $W_{13}$ wall,
\begin{equation}
W_{12}+W_{23}
\longrightarrow
W_{13}.
\end{equation}
During the collision, both radial and phase fluctuations of the Polyakov-loop field are generated in the interaction region. These excitations subsequently propagate away from the collision site, carrying away excess energy. In Fig.~\ref{su4_planar_low}, the real and imaginary parts of the Polyakov loop are shown at several stages of the evolution. 
The arrows indicate the directions of motion of the initial wall profiles. At $t=0$, the configuration consists of two domain walls, $W_{12}$ and $W_{23}$, approaching each other. At later times, these walls merge to form a single $W_{13}$ wall. Oscillatory excitations generated during the collision cause the field at the centre of the profile to fluctuate around $L=0$. This behavior is most clearly visible in the imaginary part of the Polyakov loop. At $t=152$, $Im(L)$ at the wall centre is positive, whereas at $t=177$ it
becomes negative, demonstrating the oscillatory motion of the field. 
Introducing non-zero damping suppresses these fluctuations and allows the configuration to relax to a well-defined $W_{13}$ wall.
Here, it can be clearly seen that the fluctuations in the Polyakov loop lead to fluctuations in the width of the wall.

\begin{figure}[t]
	\centering
	\includegraphics[width=0.75\linewidth]{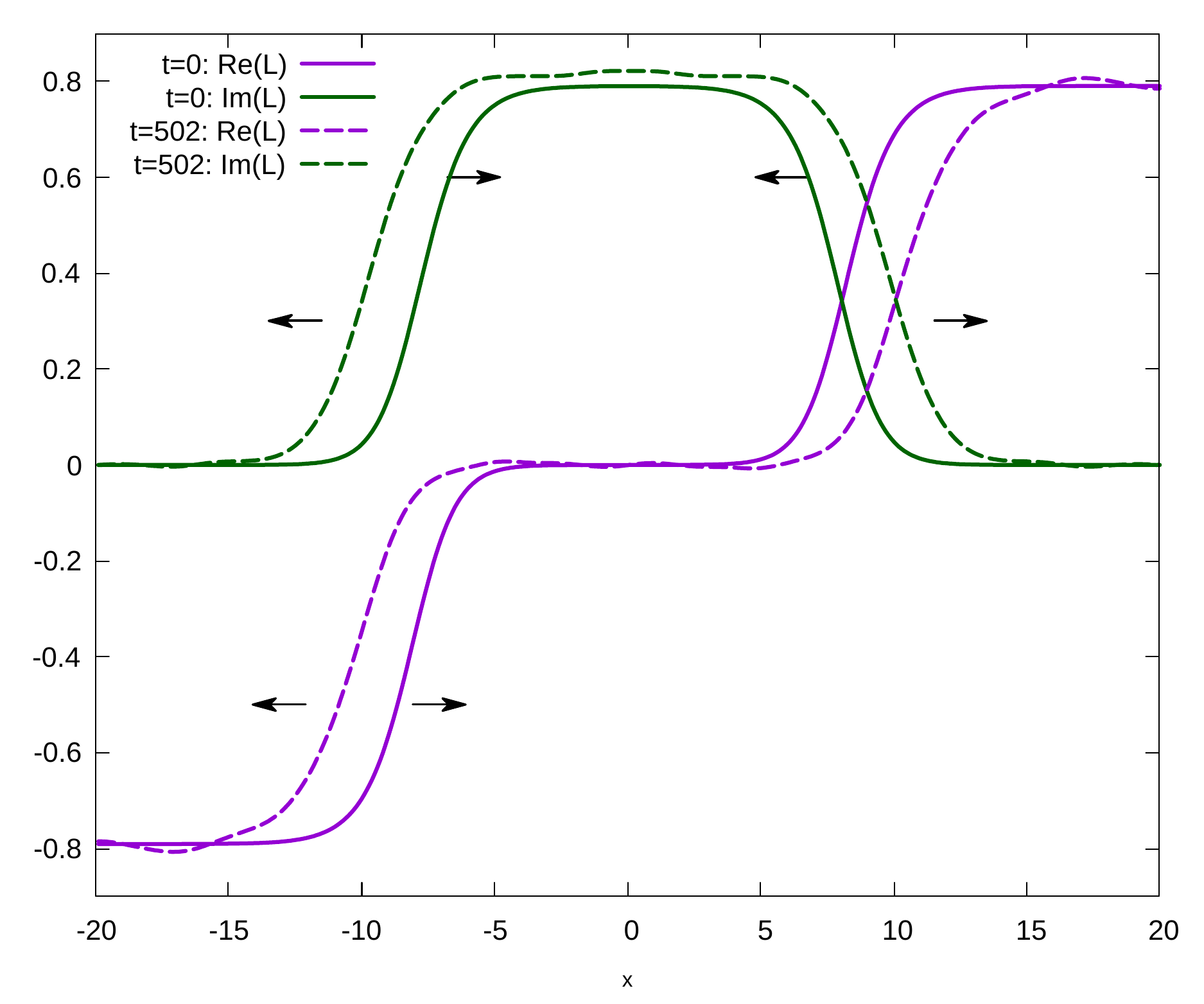}
	\caption{The evolution of real and imaginary parts of the initial (solid lines) and late-time (dashed lines) Polyakov-loop profiles for an intermediate-energy planar collision of $W_{12}$ and $W_{23}$ walls in the $Z_{4}$ theory. After temporarily forming a $W_{13}$ wall, the configuration splits back into the original $W_{12}+W_{23}$ pair, which propagates away from the collision region, as indicated by the arrows.}
	\label{su4_planar_mid}
\end{figure}

A qualitatively different behavior emerges for higher energy collisions. 
As the two walls merge, the Polyakov-loop field in the overlap region evolves from $L_2$ toward $L_4$ and subsequently returns to $L_2$.
These large oscillations drive the field away from the emerging $W_{13}$ configuration and temporarily restore a region of the $L_2$ vacuum between the walls. Consequently, the $W_{13}$ wall splits into a $W_{12}$ and $W_{23}$ pair, which subsequently move apart. The real and imaginary parts of the Polyakov loop configuration along the x-direction before and after the collision are shown in Fig.~\ref{su4_planar_mid}, with arrow marks suggesting the direction of the wall movement. 
At $t=0$, the initial $W_{12}$ and $W_{23}$ walls move towards each other.
After the collision, the two walls separate and propagate away from the collision region. 
By $t=502$, their separation is larger than that in the initial configuration.
The outgoing walls effectively retrace the trajectories of the incoming walls.

For sufficiently high collision energies, the Polyakov-loop field in the collision region overshoots the confining point $L=0$ and evolves toward the vacuum value $L_4$. As a consequence, the collision no longer results in wall fusion. Instead, the incoming walls are converted into a different pair of walls according to
\begin{equation}
W_{12}+W_{23}
\longrightarrow
W_{14}+W_{43}.
\end{equation}

\begin{figure}[t]
	\centering
	\includegraphics[width=0.65\linewidth]{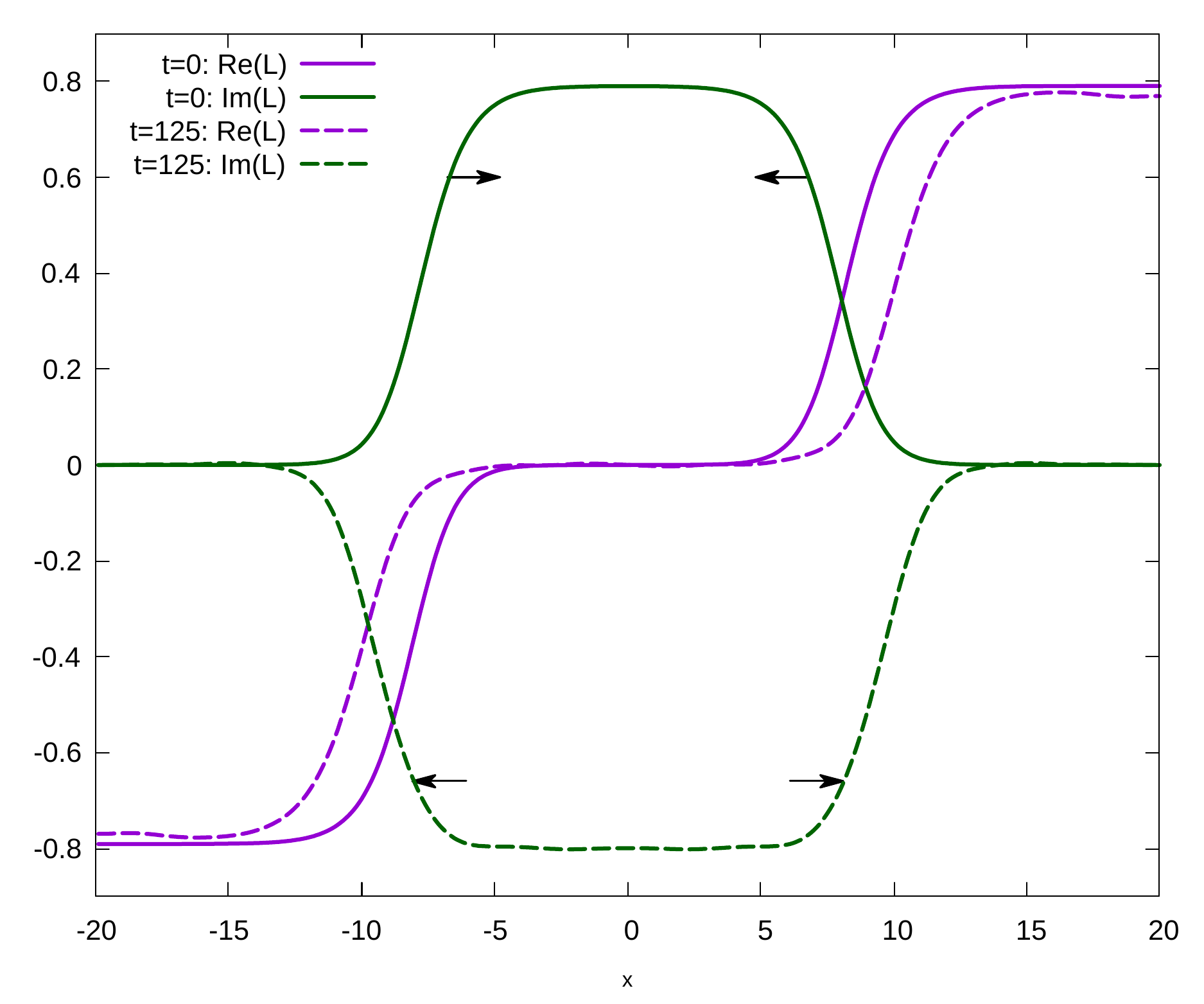}
	\caption{The evolution of real and imaginary parts of the initial (solid lines) and late-time (dashed lines) Polyakov-loop profiles for a high-energy planar collision in the $Z_{4}$ theory. The field in the collision region crosses the confining point and reaches the $L_{4}$ vacuum, converting the incoming walls through the scattering channel $W_{12}+W_{23}\rightarrow W_{14}+W_{43}$. Arrows indicate the directions of wall propagation.}
	\label{su4_planar_high}
\end{figure}

Thus, sufficiently energetic collisions open a new scattering channel that is absent in the $SU(3)$ theory. This behavior can be understood from the trajectory of the Polyakov-loop field in the complex-$L$ plane. While low-energy collisions drive the field toward the confining point $L=0$, higher collision energies allow the field to overshoot this point and reach the basin of attraction of the $L_4$ vacuum. The resulting wall configuration therefore differs qualitatively from that obtained at lower energies. A representative example of this scattering process is shown in Fig.~\ref{su4_planar_high}. 
The figure displays the real and imaginary parts of the Polyakov-loop field before and after the collision, with the arrows indicating the directions of wall motion. 
The profiles at $t=0$ represent the initial $W_{12}$ and $W_{23}$ walls. At $t=125$, the newly formed $W_{14}$ and $W_{43}$ walls move apart, and their separation exceeds
that of the initial wall pair.

More intricate phenomena arise in collisions of non-planar domain walls.  A key difference from the $SU(3)$ theory is that the low-energy fusion process
\[
W_{12}+W_{23}
\longrightarrow
W_{13},
\]
proceeds without the creation of vortex--antivortex pairs, even for non-planar collisions. Thus, wall fusion in $SU(4)$ is not intrinsically tied to the formation of topological defects.

\begin{figure}[t]
	\centering
	\includegraphics[width=0.60\linewidth]{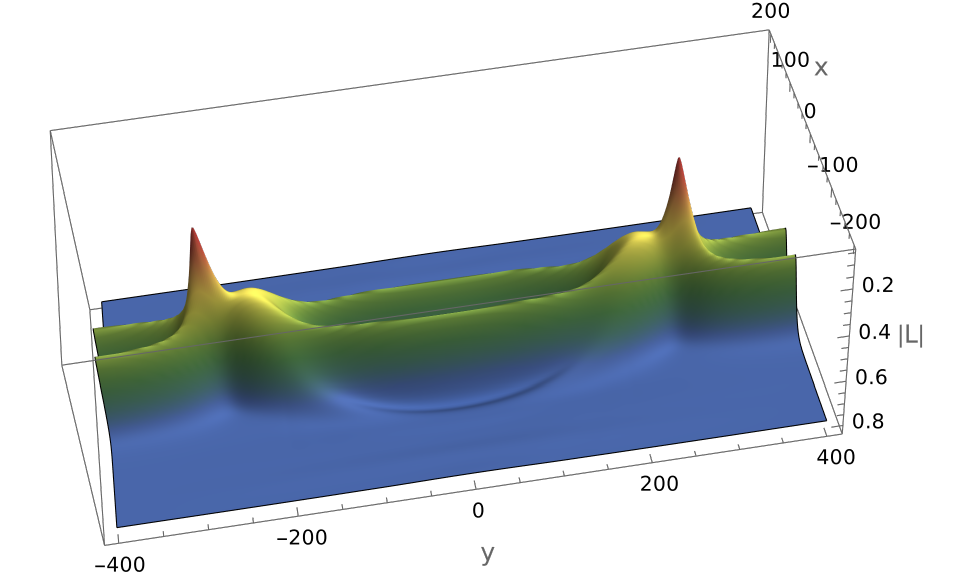}
	\caption{The formation of a vortex--antivortex pair during the collision of non-planar $W_{12}$ and $W_{23}$ domain walls in $Z_4$ theory. Spatial profiles of $|L(x,y)|$ show that the formation of vortex- antivortex appears at the endpoints of newly formed walls.}
	\label{su4_vortex_pair}
\end{figure}

For sufficiently energetic collisions, the Polyakov-loop field in the overlap region can evolve from the intermediate vacuum $L_2$ to the vacuum $L_4$, as in the planar bounce-back and transmutation processes discussed above. The subsequent evolution of this $L_4$ domain determines whether the walls ultimately transmute or retrace their original trajectories. In non-planar collisions, the resulting $L_4$ domain is bounded by $W_{14}$ and $W_{43}$ walls, whose endpoints correspond to a vortex--antivortex pair in $(2+1)$ dimensions. Depending on the collision energy and the subsequent evolution of the field, the walls may either scatter into the $W_{14}+W_{43}$ configuration or bounce back and retrace the trajectories of the incoming walls. In Fig.~\ref{su4_vortex_pair} we see the spatial profile of the Polyakov loop magnitude along the $xy-$ plane. Again, the zeros of $|L|$ correspond to the vortex and antivortex, which can be understood from the winding in the vector plot of the same magnitude profile.

This phenomenon is closely analogous to the vortex-mediated wall interactions observed in the $SU(3)$ theory. In $(3+1)$ dimensions, the vortex and antivortex are replaced by a $Z_4$ string. The wall-scattering process therefore corresponds to the nucleation of a closed string loop. As the scattering front propagates along the colliding walls, the corresponding string loop expands, providing a direct higher-dimensional interpretation of the vortex-mediated dynamics observed in $(2+1)$ dimensions.

\begin{figure}[t]
	\centering
	\includegraphics[width=0.60\linewidth]{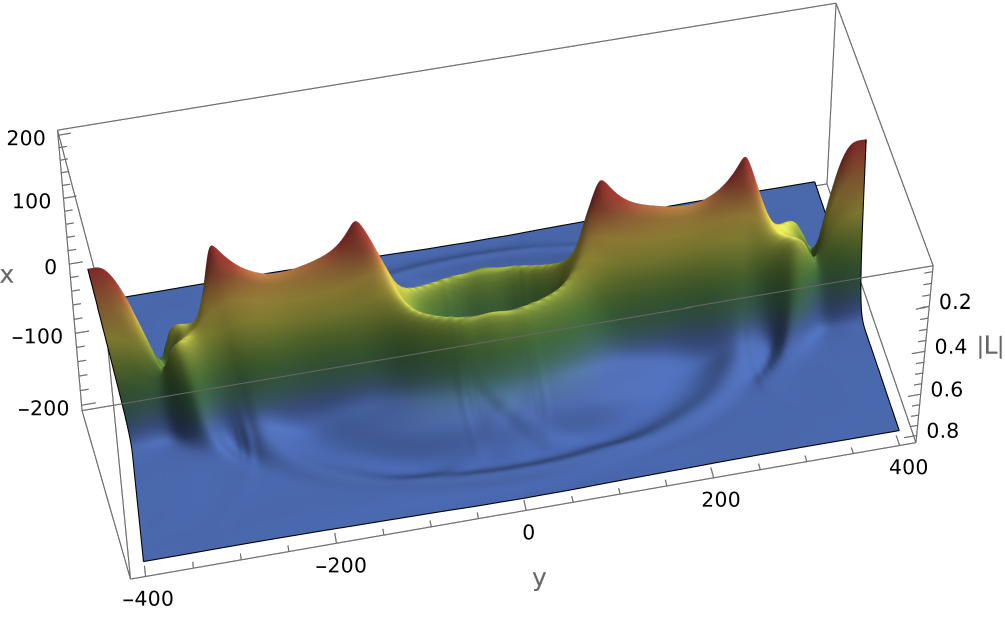}
	\caption{The formation of multiple vortex--antivortex pairs during a relativistic non-planar collision of $W_{12}$ and $W_{23}$ walls in the $Z_{4}$ theory. Large-amplitude oscillations of the Polyakov-loop field produce multiple vortex--antivortex pairs.}
	\label{su4_multiple_vav}
\end{figure}

For relativistic collisions, i.e when the wall velocities in the lab frame are $v_{x}\simeq 0.5c$, the dynamics become even richer. Large oscillations of the Polyakov-loop field are generated in the collision region, leading to repeated excursions of the field through different regions of the effective potential. As a result, additional vortex--antivortex pairs can be nucleated as observed previously in the case of $SU(3)$. As shown in Fig.~\ref{su4_multiple_vav}, the spatial profile of the Polyakov loop magnitude for the high-energy collision process leading to the excitations of the multiple vortex--antivortex pairs. The vortex--antivortex pairs are identified with the zeros of $|L|$ and from the winding of $L$ from the vector plot of the same profile.
The corresponding process in $(3+1)$ dimensions is the nucleation and subsequent collapse of additional string loops. Thus, relativistic wall collisions can produce a transient population of string loops before the system relaxes toward its final configuration.

These results demonstrate that the interactions of $SU(4)$ domain walls are qualitatively richer than those in $SU(3)$. The enlarged spectrum of wall species in the broken $Z_4$ phase leads to multiple collision outcomes, including fusion, bounce-back, and wall transmutation. While low-energy collisions are governed by the attractive interaction between neighboring walls and result in the formation of a $W_{13}$ wall, higher collision energies allow the Polyakov-loop field to access additional vacuum sectors, giving rise to qualitatively new dynamics. The interplay between domain walls and topological strings revealed by these processes highlights the rich non-equilibrium behavior of center-domain-wall networks in $SU(4)$ gauge theory.

\subsubsection{Free-energy landscape in deconfined $SU(4)$}

The low-energy collisions show that as the walls $W_{12}$ and $W_{23}$ continue to merge, the value of the Polyakov loop, $L_{mid}$, at the center between the walls continuously varies from $L_2$ to $L=0$. Thus, as in the case of $SU(3)$ we compute the free energy of a two-wall system, $F_{2W}$, in terms of $L_{mid}$.

\begin{figure}[t]
	\centering
	\includegraphics[width=0.55\linewidth]{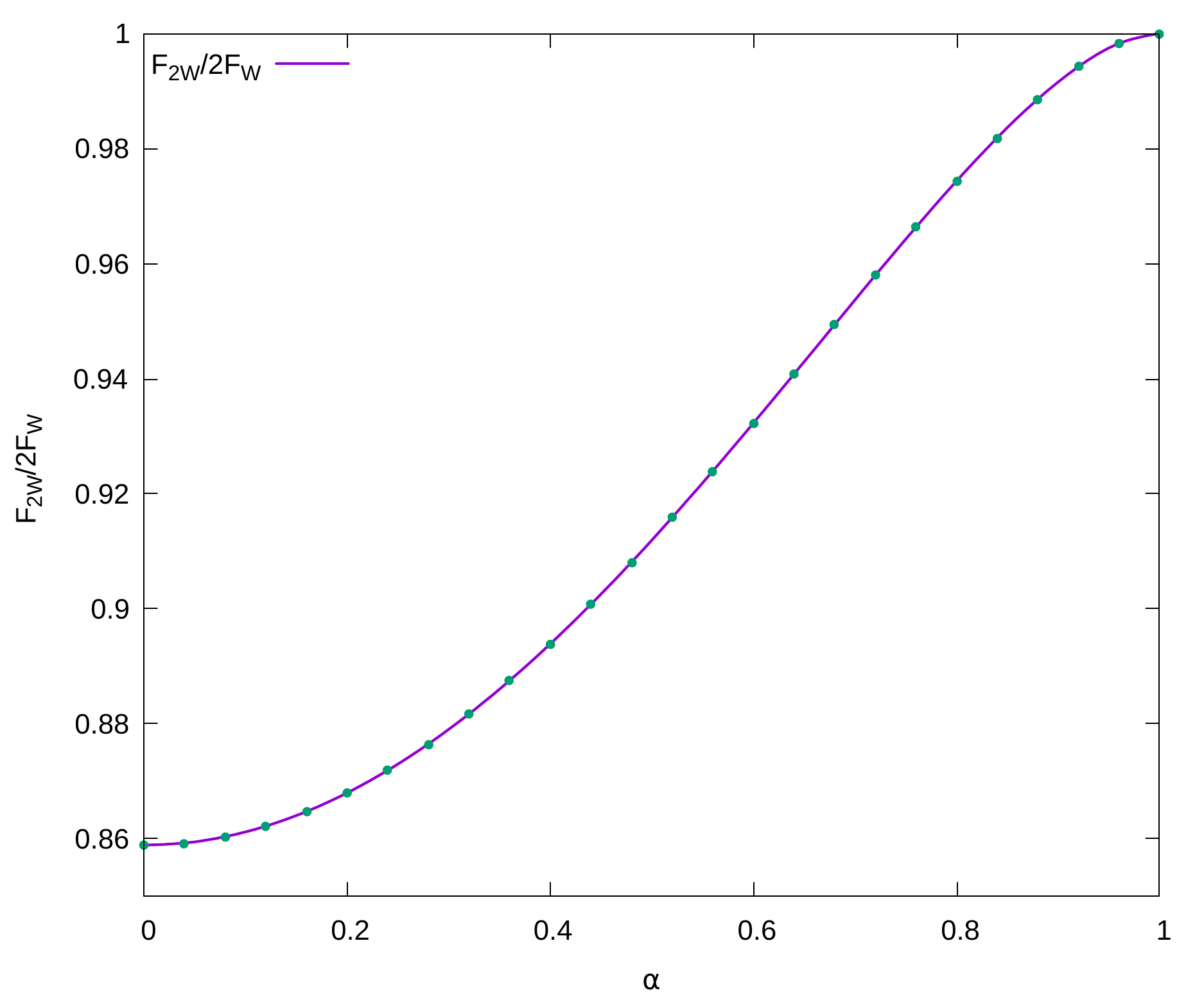}
	\caption{The normalized free energy $F_{2W}/(2F_{W})$ of a constrained $W_{12}+W_{23}$ configuration in the $Z_{4}$ theory as a function of the midpoint coordinate $\alpha$. The decrease of the free energy as the walls overlap demonstrates an attractive interaction.}
	\label{su4_free_lmid}
\end{figure}

To find the free energy of two walls at various stages of overlap, we consider a configuration consisting of a $W_{12}$ wall connecting the vacua $L_1$ and $L_2$, and a $W_{23}$ wall connecting the vacua $L_2$ and $L_3$. To quantify the interaction, we calculate the free energy of the configuration as a function of $L_{mid}$. The resulting  normalized free-energy profile is shown in Fig.~\ref{su4_free_lmid}. The free energy decreases monotonically as the midpoint value moves from $L_2$ toward $L=0$. When the midpoint field reaches $L=0$, the overlap between the two walls becomes complete, and the configuration corresponds to a single wall connecting the vacua $L_1$ and $L_3$. These results indicate that the interaction between the $W_{12}$ and $W_{23}$ walls is attractive.
The attractive interaction seen here is responsible for the large oscillations in the Polyakov loop observed even for low-energy collisions.

\section{Conclusions}

We have investigated the interaction and collision dynamics of center domain walls in the deconfined phase of $SU(N)$ gauge theories using effective Polyakov-loop models. Building on our earlier identification of topological string defects in the deconfined phase, we have explored how these defects influence the dynamics of domain-wall interactions. Our analysis combines static free-energy considerations with real-time simulations of domain-wall collisions in $2+1$ dimensions, together with their interpretation in $3+1$ dimensions.

For both $SU(3)$ and $SU(4)$ gauge theories, we find that neighboring domain walls interact attractively. The free-energy landscapes obtained from constrained two-wall configurations decrease monotonically as the walls approach one another, demonstrating that wall recombination is energetically favoured. Real-time simulations reveal the distinct dynamical mechanisms through which this energetically preferred configuration is reached.

In $SU(3)$ gauge theory, the merger of two domain walls proceeds through the creation of a vortex--antivortex pair in $2+1$ dimensions. As the vortex and antivortex separate, a segment of a third domain wall is generated between them, providing a microscopic mechanism for wall recombination. 

The $SU(4)$ theory exhibits qualitatively different behavior. In contrast to $SU(3)$, low-energy recombination of neighboring domain walls proceeds without the creation of vortex--antivortex pairs, even for non-planar collisions. Thus, wall fusion in $SU(4)$ is not intrinsically tied to the formation of topological defects. At higher collision energies, however, a new scattering channel becomes accessible in which the incoming walls are converted into a different pair of walls. In non-planar collisions, this process is accompanied by the creation of vortex--antivortex pairs, indicating that topological defects again play an essential dynamical role.

The vortex-mediated processes observed in $2+1$ dimensions admit a natural interpretation in $3+1$ dimensions. Vortices are replaced by topological strings, and vortex--antivortex pair creation corresponds to the nucleation of closed string loops. Domain-wall recombination, wall splitting, and wall scattering therefore become string-mediated processes. Our results demonstrate that topological strings are not merely static junctions of domain walls but actively participate in the non-equilibrium dynamics of defect networks.

These findings reveal a previously unexplored aspect of topological defects dynamics in deconfined non-Abelian gauge theories. The results show that topological strings are not merely static junctions connecting center domain walls but can be dynamically created during wall collisions and subsequently mediate wall recombination, wall splitting, and wall transmutation. The mechanisms identified here suggest that string-mediated interactions may be a generic feature of center-domain-wall networks in theories with spontaneously broken center symmetry. An important next step will be to investigate whether analogous processes occur in lattice simulations of the underlying gauge theories and to explore their consequences for the non-equilibrium evolution of deconfined matter near the confinement--deconfinement transition.


\begin{acknowledgments}

We thank Vinod Mamale and Indra K. Banerjee for useful comments and discussions.

\end{acknowledgments}


\end{document}